\documentclass[12pt]{article}
\usepackage{savesym}
\usepackage{cite}
\usepackage{wasysym}
\usepackage{amscd}
\usepackage{graphicx}
\usepackage{pifont}
\usepackage{float}
\usepackage{subfig}
\usepackage[all]{xy}
\usepackage{multicol}
\usepackage{amsfonts}
\usepackage{picture}
\usepackage{amssymb}
\savesymbol{iint}
\savesymbol{iiint}
\usepackage{amsmath}
\restoresymbol{TXF}{iint}
\restoresymbol{TXF}{iiint}
\usepackage{color}
\usepackage{clay}
\usepackage{hyperref}
\usepackage{slashed}
\hypersetup{colorlinks=true}
\hypersetup{linkcolor=black}
\hypersetup{citecolor=black}
\hypersetup{urlcolor=black}
\usepackage{setspace}
\usepackage{multirow}
\usepackage{wrapfig}
\usepackage{verbatim}
\numberwithin{equation}{section}
\begin{document}
\date{May, 2013}

\institution{Fellows}{\centerline{${}^{1}$Society of Fellows, Harvard University, Cambridge, MA, USA}}
\institution{HarvardU}{\centerline{${}^{2}$Jefferson Physical Laboratory, Harvard University, Cambridge, MA, USA}}

\title{Five-Dimensional Maximally Supersymmetric Yang-Mills in Supergravity Backgrounds}

\authors{Clay C\'{o}rdova,\worksat{\Fellows}\footnote{e-mail: {\tt cordova@physics.harvard.edu}}  and Daniel L. Jafferis \worksat{\HarvardU}\footnote{e-mail:{\tt jafferis@physics.harvard.edu}} }

\abstract{We determine the action for five-dimensional maximally supersymmetric Yang-Mills in off-shell supergravity backgrounds. The resulting theory contains novel five-dimensional BF type couplings as well as cubic scalar interactions which vanish in flat space. }

\maketitle

\setcounter{tocdepth}{2}
\tableofcontents
\section{Introduction}
\label{intro}

In this paper, we find the off-shell supersymmetry transformations of maximal supergravity in
five dimensions, as well as determine the (on-shell) action for a
coupled non-abelian vector multiplet. The off-shell supersymmetry
transformations and Weyl multiplet will be determined by
dimensional reduction from 6d $(2,0)$ conformal supergravity \cite{Bergshoeff:1999db, Riccioni:1997np}. Our work is the generalization to sixteen sueprcharges of \cite{Kugo:2000hn, Kugo:2000af, Fujita:2001kv}, which computed the action of 5d ${\cal N}=1$ Yang-Mills theory coupled to supergravity starting from the 6d $(1,0)$ supergravity theory \cite{Bergshoeff:1985mz, Coomans:2011ih}.  The 5d ${\cal N}=2$ Yang-Mills action coupled to the sugergravity fields originating from the 6d metric was constructed in \cite{Linander:2011jy}.  In this paper we determine the couplings of the vector multiplet to all the fields in the supergravity multiplet.

The
abelian vector multiplet can be obtained by reduction of the 6d
tensor multiplet, however the non-abelian generalization is not
known in six dimensions, so this must be determined directly by
requiring closure of the supersymmetry algebra and invariance of
the 5d action.

Most of the couplings of the 5d ${\cal N}=2$ vector multiplet to
the bosonic supergravity fields beyond covariantization of
derivatives are mass terms. However there are three more
interesting couplings, which are new interactions in the Yang-Mills multiplet induced by the background fields.  They take the following form.
\begin{itemize}
\item The graviphoton $C$ generates an interaction
\begin{equation}
\int C \wedge \mathrm{Tr}\left(\phantom{\int}\hspace{-.17in}F\wedge F\right), \label{term1} 
\end{equation}
which is the familiar Ramond-Ramond Chern-Simons term on a D4-brane in the presence of RR flux $dC$.
\item A two-form $T_{mn}$ in the $\mathbf{5}$ of the $sp(4)_{R}$ symmetry generates an interaction
\begin{equation}
\int \mathrm{Tr}\left(\phantom{\int}\hspace{-.17in}\varphi_{mn}F\right)\wedge * T^{mn}, \label{term2}
\end{equation}
where in the above $\varphi_{mn}$ are the dynamical scalars in the Yang-Mills multiplet transforming in the $\mathbf{5}$ of $sp(4)_{R}$.  The coupling \eqref{term2} is a 5d analog of a $BF$ type interaction.
\item A scalar $S_{mn}$ in the $\mathbf{10}$ of $sp(4)_{R}$ generates a cubic coupling among the vector multiplet scalars
\begin{equation}
\int S_{mn} \mathrm{Tr}\left(\phantom{\int}\hspace{-.17in}\varphi^{mr}[\varphi^{ns},\varphi_{rs}]\right). \label{term3} 
\end{equation}
The interaction \eqref{term3} is particularly novel: it vanishes in the abelian theory.  As a result this coupling has no known six-dimensional origin.  However, it is required by supersymmetry.
\end{itemize}
 
A primary application of our results occurs in the context of computing
partition functions of supersymmetric quantum field theories in
supersymmetry preserving backgrounds \cite{Pestun:2007rz, Kallen:2012cs, Hosomichi:2012ek, Kim:2012ava, Imamura:2012xg}. Such calculations yield a geometric
unification of various supersymmetric observables, and have
provided new tools in the study of strongly interacting SCFTs. One
of the insights that has been used is that the existence of
covariantly constant spinors is not necessary for preserving
supersymmetry, provided that additional appropriate operators are
activated.

The general logic is that one may couple a supersymmetric field
theory to off-shell supergravity, and fix a background of the
fields in the Weyl multiplet that is invariant under some
supersymmetries in the $M_{pl} \rightarrow 0$ limit \cite{Festuccia:2011ws}. No gravity
equation of motion, on-shell condition, or reality conditions
should be applied, since these background supergravity fields
simply keep track of the coupled terms in the dynamical quantum
field theory. The supersymmetry transformations that leave the background
configuration invariant act on the supersymmetric QFT as preserved
rigid supersymmetries.

It is our hope that the conditions \eqref{5dsusyvar} for preserving rigid supersymmetries of 5d ${\cal N}=2$ theories in general backgrounds, and the non-abelian vector multiplet action in those backgrounds \eqref{5dnonabelianaction} will prove useful in discovering new calculable supersymmetric quantities. 

\section{$6d$ $(2,0)$ Supergravity and Reduction to $5d$}
\label{offshell6dsugra}
In the section we review six-dimensional supergravity \cite{Bergshoeff:1999db}, and describe the reduction from six dimensions to five.  Specifically, the steps we take are the following.
\begin{itemize}
\item Describe the field content of off-shell $(2,0)$ supergravity.
\item Take all fields to be independent of the fifth spatial dimension, and reduce all associated representations from $so(1,5)$ to $so(1,4).$
\item Fix a gauge for the superconformal generators.  Our choice is dictated by convenience for five-dimensional calculations.
\item Identify the five-dimensional supersymmetries as six-dimensional supersymmetries combined with suitable local superconformal transformations to preserve our gauge fixing conditions.
\end{itemize}
We adopt the convention that a six-dimensional index will be underlined to distinguish it from its five-dimensional descendants.  Further conventions for spinors etc. may be found in the appendix.

Throughout our analysis we make use of the fact that our aim is to describe the coupling of quantum field theories to supergravity backgrounds.  Thus, in all supersymmetry variations and constraint equations we drop all terms proportional to fermionic fields in the supergravity multiplet. 
\subsection{$(2,0)$ Supergravity}
We begin with a brief discussion of the fields and structure of six-dimensional $(2,0)$ supergravity following \cite{Bergshoeff:1999db}.  The off-shell formulation that we utilize may be viewed as a gauge theory for the $6d$ (2,0) superconformal group.  The generators of this algebra and the associated gauge fields are given in Table \ref{6dgfields}.
\begin{table}[h]
\centering
\begin{tabular}{|c|c|c|c|c|}
\hline
Symmetry & Generators & Gauge Field & Type & Restriction \\
\hline
\multirow{2}{*}{Translations} & \multirow{2}{*}{$P_{\underline{a}}$} & \multirow{2}{*}{$e_{\underline{\mu}}^{\underline{a}}$}& \multirow{2}{*}{boson} & \multirow{2}{*}{coframe} \\
& & & & \\
\hline
\multirow{2}{*}{Lorentz} & \multirow{2}{*}{$M_{\underline{a} \underline{b}}$} & \multirow{2}{*}{$\omega_{\underline{\mu}}^{\underline{a}\underline{b}}$} & \multirow{2}{*}{boson} &\multirow{2}{*}{ spin-connection} \\
& & & & \\
\hline
$sp(4)$& \multirow{2}{*}{$U_{mn}$} & \multirow{2}{*}{$V^{mn}_{\underline{\mu}}$} & \multirow{2}{*}{boson} &\multirow{2}{*}{ $V^{mn}_{\underline{\mu}}=V^{nm}_{\underline{\mu}}$ }\\
R-symmetry & & & & \\
\hline
\multirow{2}{*}{Dilation} &\multirow{2}{*}{ $D$} & \multirow{2}{*}{$b_{\underline{\mu}}$} & \multirow{2}{*}{boson} & \multirow{2}{*}{--}\\
 & & & & \\
\hline
\multirow{2}{*}{Special Conformal} &  \multirow{2}{*}{$K_{\underline{a}}$} &  \multirow{2}{*}{$f^{\underline{a}}_{\underline{\mu}}$} &  \multirow{2}{*}{boson} &  \multirow{2}{*}{--} \\
 & & & & \\
\hline
\multirow{2}{*}{Supersymmetry} & \multirow{2}{*}{$Q_{\underline{\alpha} m}$ }& \multirow{2}{*}{$ \psi^{\underline{\alpha} m}_{\underline{\mu}}$} & \multirow{2}{*}{fermion} & $(\Gamma\psi)^{\underline{\alpha}m}_{\underline{\mu}}= \psi^{\underline{\alpha} m}_{\underline{\mu}}$ \\
 & & & & symplectic Majorana\\
 \hline
 Conformal & \multirow{2}{*}{$S_{\underline{\alpha} m}$ }& \multirow{2}{*}{$ \phi^{\underline{\alpha} m}_{\underline{\mu}}$} & \multirow{2}{*}{fermion} & $(\Gamma\phi)^{\underline{\alpha}m}_{\underline{\mu}}= \phi^{\underline{\alpha} m}_{\underline{\mu}}$ \\
 Supersymmetry& & & & symplectic Majorana\\
 \hline
\end{tabular}
\caption{Gauge fields of six-dimensional conformal supergravity}
\label{6dgfields}
\end{table}

Although the structure of the supergravity multiplet can be understood from the superconformal group, an important feature is that not all of the fields appearing in Table \ref{6dgfields} are independent.  There are constraints which relate $\omega_{\underline{\mu}}^{\underline{ab}}, f_{\underline{\mu}}^{\underline{a}},$ and $ \phi^{\underline{\alpha} i}_{\underline{\mu}}$ to other fields in the multiplet.  We have need only of the following
\begin{eqnarray}
\omega_{\underline{\mu}}^{\underline{ab}}  & = &  2e^{\underline{\nu}[\underline{a}}\partial_{[\underline{\mu}}e_{\underline{\nu}]}^{\hspace{.1in}\underline{b}]} -e^{\underline{\rho}[\underline{a}}e^{\underline{b}]\underline{\sigma}}e_{\underline{\mu}}^{\underline{c}}\partial_{\underline{\rho}}e_{\underline{\sigma c}}+2e_{\underline{\mu}}^{[\underline{a}}b^{\underline{b}]}\label{spinconst}\\
f_{\underline{a}}^{\underline{a}} & = & \frac{1}{20}\underline{R}, \nonumber
\end{eqnarray}
where, in the above, $\underline{R}$ is the scalar curvature of the connection $\omega_{\underline{\mu}}^{\underline{ab}}$.   In the special case where the dilation gauge field $b_{\underline{\mu}}$ vanishes, $\omega_{\underline{\mu}}^{\underline{ab}}$ is the spin connection of the six-dimensional metric and $\underline{R}$ is the ordinary Ricci scalar curvature, however in a general background both quantities receive corrections.

Another important feature of this supergravity is that off-shell closure of the supersymmetry algebra can only be achieved provided that additional auxiliary fields are included beyond the gauge fields indicated in Table \ref{6dgfields}.  The necessary fields are indicated in Table \ref{6dmfields} below.
\begin{table}[h]
\centering
\begin{tabular}{|c|c|c|c|c|}
\hline
Field & Type & Restriction & $sp(4)_R $ & Weyl Weight \\
\hline
\multirow{2}{*}{$ T^{mn}_{\underline{abc}}$} & \multirow{2}{*}{boson} &$T^{mn}_{\underline{abc}}=-\frac{1}{6}\varepsilon_{\underline{abc}}^{\phantom{\underline{abc^{d}}}\hspace{-.03in}\underline{def}}T^{mn}_{\underline{def}},$ & \multirow{2}{*}{$\mathbf{5}$} & \multirow{2}{*}{1} \\
 & & $T^{mn}_{\underline{abc}}=-\phantom{\frac{1}{1_{a_{a_{a}}}}}\hspace{-.27in}T^{nm}_{\underline{abc}}, \hspace{.2in}\phantom{A^{A^{A^{A}}}}\Omega_{mn}T^{mn}_{\underline{abc}}=0.$ & &\\
\hline
\multirow{2}{*}{$ D^{mn,rs}$} & \multirow{2}{*}{boson} & $D^{mn,rs}=D^{rs,mn}=-D^{nm,rs}=-D^{mn,sr},$& \multirow{2}{*}{$\mathbf{14}$} & \multirow{2}{*}{2} \\
 & & $\Omega_{mn}D^{mn,rs}=\Omega_{rs}D^{mn,rs}=\Omega_{mr}\Omega_{ns}D^{mn,rs}=0.$ & &\\
\hline
\multirow{2}{*}{$ \chi^{\underline{\alpha}mn }_{r}$} &\multirow{2}{*}{fermion} & $\chi^{mn}_{r}=-\chi^{nm}_{r}, \hspace{.5in} \Omega_{mn}\chi^{mn}_{r}=\delta^{r}_{m}\chi^{mn}_{r}=0,$& \multirow{2}{*}{$\mathbf{16}$} & \multirow{2}{*}{3/2} \\
& & $\left(\Gamma \chi \right)^{\underline{\alpha}mn }_{r}=\chi^{\underline{\alpha}mn }_{r}, \hspace{.5in} \mathrm{symplectic \  Majorana.}$ & & \\
\hline
\end{tabular}
\caption{Matter fields of six-dimensional conformal supergravity}
\label{6dmfields}
\end{table}

Given these fields one may now write down the supersymmetry transformations which depend on local symplectic Majorana-Weyl Grassmann parameters $\epsilon^{m}$, and $\eta^{m}$, of positive and negative chirality respectively, associated to local supersymmetry and local conformal supersymmetry transformations.  The only explicit forms we require are
\begin{eqnarray}
\delta e_{\underline{\mu}}^{\underline{a}} & = & \frac{1}{2}\overline{\epsilon}\Gamma^{\underline{a}}\psi_{\underline{\mu}}, \label{6dsusy}\\
\delta b_{\underline{\mu}} & = & -\frac{1}{2}\overline{\epsilon} \phi_{\underline{\mu}}+\frac{1}{2} \overline{\eta}\psi_{\underline{\mu}}, \nonumber \\
\delta \psi_{\underline{\mu}}^{m} & = & \mathcal{D}_{\underline{\mu}}\epsilon^{m}+\frac{1}{24}T^{mn}_{\underline{abc}}\Gamma^{\underline{abc}}\Gamma_{\underline{\mu}}\epsilon_{n}+\Gamma_{\underline{\mu}}\eta^{m}, \nonumber \\
\delta \chi_{r}^{mn} & = & \frac{5}{32}\left(\mathcal{D}_{\underline{\mu}}T^{mn}_{\underline{abc}}\right)\Gamma^{\underline{abc}}\Gamma^{\underline{\mu}}\epsilon_{r}-\frac{15}{16}\Gamma^{\underline{\mu \nu}}R_{\underline{\mu \nu}r}^{[m}\epsilon^{n]}-\frac{1}{4}D^{mn}_{rs}\epsilon^{s}+\frac{5}{8}T^{mn}_{\underline{abc}}\Gamma^{\underline{abc}}\eta_{r}-(\mathrm{traces}). \nonumber
\end{eqnarray}
where in the above, the covariant derivatives and $sp(4)$ curvature tensor $R_{\underline{\mu \nu}}^{mn}$ are
\begin{eqnarray}
\mathcal{D}_{\underline{\mu}}\epsilon^{m} & = & \partial_{\underline{\mu}}\epsilon^{m}+\frac{1}{2}b_{\underline{\mu}}\epsilon^{m}+\frac{1}{4}\omega_{\underline{\mu}}^{\underline{ab}}\Gamma_{\underline{ab}}\epsilon^{m}-\frac{1}{2}V_{\underline{\mu}n}^{m}\epsilon^{n}, \\
\mathcal{D}_{\underline{\mu}}T^{mn}_{\underline{abc}} & = & \partial_{\underline{\mu}}T^{mn}_{\underline{abc}}+3\omega_{\underline{\mu}[\underline{a}}^{\underline{d}}T^{mn}_{\underline{bc}] \underline{d}}-b_{\underline{\mu}}T^{mn}_{\underline{abc}}+V_{\underline{\mu}r}^{[m}T^{n]r}_{\underline{abc}}-(\mathrm{traces}), \nonumber \\
R_{\underline{\mu\nu}}^{mn} & = & 2\partial_{[\underline{\mu}}V^{mn}_{\underline{\nu}]}+V_{[\underline{\mu}}^{r(m}V_{\underline{\nu}]r}^{n)}, \nonumber
\end{eqnarray}
and the notation ``traces" indicates terms proportional to $sp(4)$ invariant tensors $\Omega_{mn},$ and $\delta^{m}_{n}$.

Finally, we also have need of the variation of the independent gauge fields under bosonic gauge transformations.  Let $\Lambda_{D}, \Lambda_{K\underline{a}}, \Lambda^{\underline{a b}},$ and $\Lambda^{mn}$ denote the parameters of infinitesimal dilation, special conformal, Lorentz, and $sp(4)$ transformations.  Then the variations are 
\begin{eqnarray}
\delta e_{\underline{\mu}}^{\underline{a}} & = & -\Lambda_{D}e_{\underline{\mu}}^{\underline{a}}-\Lambda^{\underline{ab}}e_{\underline{\mu b}}, \label{6dgaugevar}\\
\delta V_{\underline{\mu}}^{mn} & = & \partial_{\underline{\mu}}\Lambda^{mn}+\Lambda^{(m}_{r}V^{n)r}_{\underline{\mu}}, \nonumber \\
\delta b_{\underline{\mu}} & = & \partial_{\underline{\mu}}\Lambda_{D}-2e^{\underline{a}}_{\underline{\mu}}\Lambda_{K \underline{a}}, \nonumber\\
\delta \psi_{\underline{\mu}}^{m} & = & -\frac{1}{2}\Lambda_{D}\psi^{m}_{\underline{\mu}} +\frac{1}{2}\Lambda^{m}_{n}\psi^{n}_{\underline{\mu}}-\frac{1}{4}\Lambda^{\underline{ab}}\Gamma_{\underline{ab}}\psi^{m}_{\underline{\mu}}.\nonumber
\end{eqnarray}
The transformations of the additional auxiliary fields follow from their representations indicated in Table \ref{6dmfields}.  In particular, we note that all matter fields are neutral under special conformal transformations and any field $\Theta$ of Weyl weight $w$ transforms under dilatation as
\begin{equation}
\delta_{D} \Theta=w \Lambda_{D}\Theta.
\end{equation}
Thus, $w$ labels the scaling dimensions of fields.

\subsection{Reduction to Five-Dimensions}
Now we proceed with dimensional reduction following the logic of \cite{Kugo:2000hn}.  We take all fields to be independent of the direction with vector index $\underline{\mu}=z$ with associated frame index $\underline{a}=5$, and decompose the frame and coframe as  
\begin{equation}
e^{\underline{\mu}}_{\underline{a}}=\left(\begin{array}{cc}e^{\mu}_{a} & e^{z}_{a}=-C_{a} \\
e^{\mu}_{5}=0 & e^{z}_{5}=\alpha
\end{array} \right), \hspace{.5in} e^{\underline{a}}_{\underline{\mu}}=\left(\begin{array}{cc}e^{a}_{\mu} & e^{5}_{\mu}=\alpha^{-1}C_{\mu} \\
e^{a}_{z}=0 & e^{5}_{z}=\alpha^{-1}
\end{array} \right). \label{coframered}
\end{equation}
The field $C_{\mu}$ appearing in the above is a five-dimensional gauge field referred to as the graviphoton.  The scalar $\alpha$ is the dilaton.\footnote{Often the field $\alpha$ is rewritten as $\alpha =e^{\sigma}$ and $\sigma$ is referred to as the dilation.  However, as our interest is in supergravity backgrounds, there is no relevant notion of a canonically normalized Einstein-Hilbert action and hence we find it more convenient to deal directly with the field $\alpha$ and utilize the naming convention stated above. }  

The decomposition appearing in \eqref{coframered} requires a partial gauge fixing of the local Lorentz group.  We find it convenient to fix other generators via the following conditions 
\begin{equation}
M_{a5}: \ e^{a}_{z}=0, \hspace{.5in} S_{\underline{\alpha} m}:  \ \psi^{\underline{\alpha} m}_{5}=0, \hspace{.5in} K_{a}:  b_{\mu}-\alpha^{-1}\partial_{\mu}\alpha=0, \hspace{.5in}K_{5}: \ b_{5}=0.  \label{gauge}
\end{equation}
This choice has the advantage that the form of the frame is invariant under supersymmetry transformations.  

The decomposition of the coframe \eqref{coframered} specifies how the six-dimensional metric descends to a five-dimensional fields.  The remaining bosons similarly are reduced as
\begin{eqnarray}
V_{\underline{a}}^{mn} & \rightarrow &\begin{cases} V_{a}^{mn} & \underline{a}\neq 5  \\
V_{5}^{mn}\equiv S^{mn} & \underline{a} = 5\end{cases}, \nonumber\\
T^{mn}_{\underline{abc}} & \rightarrow & T^{mn}_{ab5} \equiv T^{mn}_{ab},\\
D^{mn,rs} & \rightarrow & D^{mn,rs}. \nonumber
\end{eqnarray}
The fields on the right-hand-side are the independent five-dimensional bosons needed for maximally supersymmetric off-shell supergravity.  They are related to the six-dimensional parent fields by matching of tangent indices (as opposed to coordinate indices).  Such a matching ensures that the reduced fields transform appropriately under five-dimensional coordinate transformations.  Also, we note that the anti-self-duality constraint on $T^{mn}_{\underline{abc}}$ implies that its independent five-dimensional descendants consist of a collection of two-forms in five dimensions.

We may similarly decompose the six-dimensional fermions.  As described in detail in section \ref{6d5dspinors}, the eight component Dirac spinor of $so(1,5)$ may be viewed as a doublet of four component Dirac spinors of $so(1,4)$.  Making use of the gauge fixing conditions \eqref{gauge} on the gravitino, we have\footnote{Numerical prefactors in these formulas are chosen for later convenience. }
\begin{eqnarray}
\psi_{\underline{a}}^{\underline{\alpha}m} & \rightarrow & \left(\begin{array}{c} \psi_{a}^{\alpha m} \\ 0\end{array}\right),\\
\chi^{\underline{\alpha}mn}_{r} & \rightarrow &\frac{15}{16}  \left(\begin{array}{c} \chi^{\alpha mn}_{r} \\ 0\end{array}\right). \nonumber
\end{eqnarray}
The fields on the right-hand-side are the independent five-dimensional fermions needed for maximally supersymmetric off-shell supergravity.  They are symplectic Majorana in the five-dimensional sense.  In the conventions set in appendix \ref{5dcliffcon} this means that these spinors are equal to their Dirac conjugates
\begin{equation}
\overline{\psi}_{a}^{\alpha m} = \psi_{a}^{\alpha m}, \hspace{.5in}\overline{\chi}^{\alpha mn}_{r} =\chi^{\alpha mn}_{r}.
\end{equation}
A complete summary of all the five-dimensional supergravity fields is given in Table \ref{5dgravfields}.
\begin{table}[h]
\centering
\begin{tabular}{|c|c|c|c|c|}
\hline
Field & Type & Restriction & $sp(4)_R $ & Weyl Weight \\
\hline
\multirow{2}{*}{$e^{a}_{\mu}$} & \multirow{2}{*}{boson} & \multirow{2}{*}{coframe} &\multirow{2}{*}{$\mathbf{1}$} & \multirow{2}{*}{-1}\\
 & &  & & \\
 \hline
 \multirow{2}{*}{$C_{\mu}$} & \multirow{2}{*}{boson} & \multirow{2}{*}{``graviphoton''  $ G\equiv dC$} &\multirow{2}{*}{$\mathbf{1}$} & \multirow{2}{*}{0}\\
 & &  & & \\
\hline
 \multirow{2}{*}{$\alpha$} & \multirow{2}{*}{boson} & \multirow{2}{*}{``dilaton''} &\multirow{2}{*}{$\mathbf{1}$} & \multirow{2}{*}{1}\\
 & &  & & \\
\hline
 \multirow{2}{*}{$V_{\mu}^{mn}$} & \multirow{2}{*}{boson} & \multirow{2}{*}{$V_{\mu}^{mn}=V_{\mu}^{nm}$} &\multirow{2}{*}{$\mathbf{10}$} & \multirow{2}{*}{0}\\
 & &  & & \\
\hline
\multirow{2}{*}{$S^{mn}$} & \multirow{2}{*}{boson} & \multirow{2}{*}{$S^{mn}=S^{nm}$} &\multirow{2}{*}{$\mathbf{10}$} & \multirow{2}{*}{1}\\
 & &  & & \\
\hline
\multirow{2}{*}{$\psi^{\alpha m}_{\mu}$} & \multirow{2}{*}{fermion} & \multirow{2}{*}{symplectic Majorana ``gravitini''} &\multirow{2}{*}{$\mathbf{4}$} & \multirow{2}{*}{$-1/2$}\\
 & &  & & \\
\hline
\multirow{2}{*}{$ T^{mn}_{\mu \nu}$} & \multirow{2}{*}{boson} &$T^{mn}_{\mu \nu}=-T^{mn}_{\nu \mu},$ & \multirow{2}{*}{$\mathbf{5}$} & \multirow{2}{*}{-1} \\
 & & $T^{mn}_{\mu \nu}=-\phantom{\frac{1}{1_{a_{a_{a}}}}}\hspace{-.27in}T^{nm}_{\mu \nu}, \hspace{.2in}\phantom{A^{A^{A^{A}}}}\Omega_{mn}T^{mn}_{\mu \nu}=0.$ & &\\
\hline
\multirow{2}{*}{$ D^{mn,rs}$} & \multirow{2}{*}{boson} & $D^{mn,rs}=D^{rs,mn}=-D^{nm,rs}=-D^{mn,sr},$& \multirow{2}{*}{$\mathbf{14}$} & \multirow{2}{*}{2} \\
 & & $\Omega_{mn}D^{mn,rs}=\Omega_{rs}D^{mn,rs}=\Omega_{mr}\Omega_{ns}D^{mn,rs}=0.$ & &\\
\hline
\multirow{2}{*}{$ \chi^{\alpha mn }_{r}$} &\multirow{2}{*}{fermion} & $\chi^{mn}_{r}=-\chi^{nm}_{r}, \hspace{.5in} \Omega_{mn}\chi^{mn}_{r}=\delta^{r}_{m}\chi^{mn}_{r}=0,$& \multirow{2}{*}{$\mathbf{16}$} & \multirow{2}{*}{3/2} \\
& & symplectic Majorana  ``dilatini'' & & \\
\hline
\end{tabular}
\caption{Fields of five-dimensional off-shell $\mathcal{N}=2$ supergravity.}
\label{5dgravfields}
\end{table}
For the remainder of this paper, all supergravity calculations will be carried out in five dimensions using these fields.
\subsection{Identification of $5d$ Supersymmetry}

Our next task is to reduce the supersymmetry transformations from six to five dimensions.   A general variation \eqref{6dsusy} does not respect the gauge fixing conditions \eqref{gauge}, and hence does not respect our identification of the five-dimensional fields.  To remedy this we construct a five-dimensional supersymmetry variation, $\delta_{Q}(\epsilon),$ preserving the gauge fixing conditions.  Such a transformation takes the general form 
\begin{equation}
\delta_{Q}(\epsilon)=\delta_{\underline{Q}}(\epsilon)+\delta_{\underline{S}}(\eta(\epsilon))+\delta_{\underline{K}}(\Lambda_{K}(\epsilon)).
\end{equation}
On the right-hand-side of the above, appears six-dimensional transformations associated to supersymmetry, conformal supersymmetry, and special conformal transformations.  Such terms correct the six-dimensional supersymmetry transformations and are uniquely fixed by requiring the five-dimensional supersymmetry to preserve \eqref{gauge}.  

In our context, we are only interested in determining which backgrounds are supersymmetric in five dimensions, and hence we only require the variation of the fermions.  These variations are determined once $\eta$ is known as a function of $\epsilon$.  This relationship may be fixed by varying the gauge condition on the gravitino.  We carry out this calculation in five-dimensional notation where (as with the fermions of the previous section), the spinor parameters reduce as
\begin{equation}
\epsilon^{m}_{\underline{\alpha}}\rightarrow \left(\begin{array}{c} \epsilon^{m}_{\alpha}   \\ 0\end{array}\right), \hspace{.5in} \eta^{m}_{\underline{\alpha}}\rightarrow \left(\begin{array}{c}   0 \\ i\eta^{m}_{\alpha}\end{array}\right).
\end{equation}
Then, the variation of the gravitino constraint is
\begin{equation}
\delta_{Q}\left(\psi_{5}^{m}\right) =  -\frac{1}{8\alpha}G^{ab}\Gamma_{ab}\epsilon^{m}-\frac{1}{2}S^{m}_{n}\epsilon^{n}+\frac{1}{4}T^{mn}_{ab}\Gamma^{ab}\epsilon_{n}+\eta^{m}.\label{gravitinovar65}
\end{equation}
In the above, $G_{ab}$ is the graviphoton field strength, and in simplifying \eqref{gravitinovar65} we have made use of the following facts.
\begin{itemize}
\item In the gauge \eqref{gauge}, the components of the generalized spin connection involving the reduction dimension are
\begin{equation}
\omega_{z}^{ab}=-\frac{1}{2\alpha^{2}}G^{ab},\hspace{.5in}\omega_{\mu}^{a5}=\frac{1}{2\alpha}e^{\nu a}G_{\mu\nu}, \hspace{.5in} \omega_{z}^{a5}=0.
\end{equation}
\item The anti-self-dual three-form reduces as
\begin{equation}
T^{mn}_{\underline{abc}}\Gamma^{\underline{abc}}=-6iT^{mn}_{ab}\Gamma^{ab}\otimes \left(\begin{array}{cc}0 & 1 \\ 0 & 0\end{array}\right).
\end{equation}
\end{itemize}

Demanding that \eqref{gravitinovar65} vanish determines $\eta$, and allows us the write the explicit form of the five-dimensional supersymmetry variations of the fermions by simplifying  \eqref{6dsusy}.  The result is
\begin{eqnarray}
\delta \psi_{a}^{m} & = & \mathcal{D}_{a}\epsilon^{m}+\frac{i}{2\alpha}\left[\phantom{\frac{1}{1}}\hspace{-.11in}G_{ab}\Omega^{mn}-\alpha S^{mn}\eta_{ab}\right]\Gamma^{b}\epsilon_{n}+\frac{i}{8\alpha}\left[\phantom{\frac{1}{1}}\hspace{-.11in}G^{bc}\Omega^{mn}-4\alpha\left(T^{mn}\right)^{bc}\right]\Gamma_{abc}\epsilon_{n}, \nonumber \\
\delta \chi^{mn}_{r} & = &\left[\phantom{\frac{1}{1}}\hspace{-.1in}T^{mn}_{ab}T_{cdrs}-\frac{1}{\alpha}T^{mn}_{ab}G_{cd}\Omega_{rs}+\frac{1}{12}\left(\mathcal{D}^{e}S^{[m}_{r}\delta^{n]}_{s}+\mathcal{D}_{f}T^{mn f e}\Omega_{rs}\right)\varepsilon_{eabcd}\right]\Gamma^{abcd}\epsilon^{s} \nonumber \\
& + &\left[ \frac{5}{2\alpha}T^{mn}_{ab}G^{a}_{\phantom{a}c}\Omega_{rs}-4 T^{mn}_{ab}T^{a}_{\phantom{a}crs}+2T^{mn}_{bc}S_{rs}- S_{p}^{[m}T^{n]p}_{bc}\Omega_{rs}-R_{bc r}^{\phantom{b}[m}\delta^{n]}_{s}\right.  \label{5dsusyvar} \\
& + & \left.\frac{1}{2}\mathcal{D}_{a}T^{mn}_{de}\Omega_{rs}\varepsilon^{ade}_{\phantom{ade}bc}\right]\Gamma^{bc}\epsilon^{s}+\left[\frac{1}{\alpha}T^{mn}_{ab}G^{ab}\Omega_{rs}-2 T^{mn}_{ab}T^{ab}_{rs}-\frac{4}{15}D^{mn}_{rs}\right]\epsilon^{s}  - (\mathrm{traces}). \nonumber
\end{eqnarray}
The five-dimensional covariant derivatives, curvatures, and connections are
\begin{eqnarray}
\mathcal{D}_{\mu}\epsilon^{m} & = & \partial_{\mu}\epsilon^{m}+\frac{1}{2}\partial_{\mu}\log(\alpha)\epsilon^{m}+\frac{1}{4}\omega_{\mu}^{bc}\Gamma_{bc}\epsilon^{m}-\frac{1}{2}V_{\mu n}^{m}\epsilon^{n},  \nonumber\\
\mathcal{D}_{\mu}S^{m n} & = & \partial _{\mu}S^{mn}-\partial_{\mu}\log(\alpha)S^{mn}-V_{\mu r}^{(m}S^{n)r}, \nonumber \\
\mathcal{D}_{\mu}T^{mn}_{ab} & = & \partial_{\mu}T^{mn}_{ab}-2\omega_{\mu [a}^{c}T^{mn}_{b]c}-\partial_{\mu}\log(\alpha)T^{mn}_{ab} + V_{\mu s}^{[m}T^{n]s}_{ab}, \label{5dcurvdefs}\\
R_{\mu \nu}^{mn} & = & 2\partial_{[\mu}V^{mn}_{\nu]}+V_{[\mu}^{r(m}V_{\nu]r}^{n)}, \nonumber \\
\omega_{\mu}^{ab} & = &  2e^{\nu[a}\partial_{[\mu}e_{\nu]}^{\hspace{.1in}b]} -e^{\rho[a}e^{b]\sigma}e_{\mu}^{c}\partial_{\rho}e_{\sigma c}+2e_{\mu}^{[a}\partial^{b]}\log(\alpha). \nonumber
\end{eqnarray}

Equations \eqref{5dsusyvar} are the key results of this section.  A supersymmetric background of a five-dimensional field theory coupled to supergravity is one for which there exists a nowhere vanishing spinor $\epsilon^{m}$ such that the variations $\delta \psi_{a}^{m}$ and $\delta \chi^{mn}_{r}$ vanish.  In the special case where $\alpha$ is constant and all background fields other than the five-dimensional metric are turned off, \eqref{5dsusyvar} reduces to the killing spinor equation on $\epsilon^{m}$.  However, if we activate more general values of the supergravity background fields, the conditions for preserving supersymmetry are much less constraining.

\section{Tensor Multiplets and Reduction to $5d$ Yang-Mills}
\label{5dtensred}
In this section we extend our previous analysis by coupling a $(2,0)$ tensor multiplet to off-shell supergravity.  By reducing to five dimensions, we obtain an action for maximally supersymmetric five-dimensional abelian Yang-Mills coupled to background supergravity fields.  The extension of this action to non-abelian Yang-Mills is carried out directly in five dimensions and described in section \ref{nonabext}.  

\subsection{Tensor Multiplets in Supergravity Backgrounds}
We begin with the tensor multiplet in six-dimensions in the absence of supergravity background fields. The fields of this multiplet are enumerated in Table \ref{6dtensorfields}.  
\begin{table}[h]
\centering
\begin{tabular}{|c|c|c|c|c|}
\hline
Field & Type & Restriction & $sp(4) _R$ & Weyl Weight \\
\hline
\multirow{2}{*}{$B_{\underline{\mu \nu}}$} & \multirow{2}{*}{boson} & \multirow{2}{*}{Chiral two-form gauge field.} &\multirow{2}{*}{$\mathbf{1}$} & \multirow{2}{*}{0}\\
 & &  & & \\
 \hline
 \multirow{2}{*}{$\Phi^{mn}$} & \multirow{2}{*}{boson} & \multirow{2}{*}{$\Phi^{mn}=-\Phi^{nm}, \hspace{.5in}\Omega_{mn}\Phi^{mn}=0.$} &\multirow{2}{*}{$\mathbf{5}$} & \multirow{2}{*}{2}\\
 & &  & & \\
\hline
 \multirow{2}{*}{$\varrho^{\underline{\alpha}m}$} & \multirow{2}{*}{fermion} & \multirow{2}{*}{$\left(\Gamma \varrho\right)^{\underline{\alpha}m}=-\varrho^{\underline{\alpha}m},$ \hspace{.5in}symplectic Majorana.} &\multirow{2}{*}{$\mathbf{4}$} & \multirow{2}{*}{5/2}\\
 & &  & & \\
\hline
\end{tabular}
\caption{Fields of the six-dimensional on-shell tensor multiplet.}
\label{6dtensorfields}
\end{table}

The two-form $B$ is a gauge field meaning that its value may be shifted by an arbitrary exact two-form $d\Lambda$.  Further, it is chiral meaning that its gauge invariant field strength $H=dB$ is self-dual.  In components this constraint takes the form
\begin{equation}
H_{\underline{abc}}=\frac{1}{6}\varepsilon_{\underline{abcdef}}H^{\underline{def}}.
\end{equation}
In general in the following we use the superscript $\pm$ on a three-form $\Xi$ to indicate the self-dual and anti-self-dual projections of the form
\begin{equation}
\Xi^{+}\equiv \frac{1}{2}\left(\Xi+*\Xi\right), \hspace{.5in}\Xi^{-}\equiv \frac{1}{2}\left(\Xi-*\Xi\right).
\end{equation}

Unlike the supergravity multiplets discussed in previous sections, the tensor multiplet is an on-shell multiplet meaning that the match between bosonic and fermionic degrees of freedom is only achieved after the equations of motion are enforced.  In the flat supergravity background these take the form of free field equations of motion
\begin{equation}
dH=0, \hspace{.5in}\partial^{2}\Phi^{mn}=0, \hspace{.5in} \slashed{\partial}\varrho^{m}=0.
\end{equation}

Now let us describe the coupling of these fields to off-shell supergravity backgrounds.  The general conformal supersymmetry variation is given by
\begin{eqnarray}
\delta B_{\underline{\mu\nu}} & = & \overline{\epsilon}_{m}\Gamma_{\underline{\mu\nu}}\varrho^{m},  \nonumber \\
\delta \Phi^{mn} & = &- 4\overline{\epsilon}^{[m}\varrho^{n]}-\Omega^{mn}\overline{\epsilon}^{r}\varrho_{r}, \label{6dtensorvar}\\
\delta \varrho^{m} & = & \frac{1}{48}H^{+}_{\underline{\mu\nu\sigma}}\Gamma^{\underline{\mu \nu \sigma}}\epsilon^{m} +\frac{1}{4}\slashed{\mathcal{D}}\Phi^{mn}\epsilon_{n}-\Phi^{mn}\eta_{n} . \nonumber
\end{eqnarray}
These variations are consistent provided that the following equations of motion are satisfied
\begin{eqnarray}
H^{-}_{\underline{\mu\nu \sigma}}-\frac{1}{2}\Phi_{mn}T^{mn}_{\underline{\mu\nu \sigma}}& = &0, \nonumber \\
\mathcal{D}^{2}\Phi_{mn}-\frac{1}{15}D^{rs}_{mn}\Phi_{rs}+\frac{1}{3}H^{+}_{\underline{\mu\nu \sigma}}T_{mn}^{\underline{\mu\nu \sigma}} & = & 0, \label{6dtensoreom} \\
\slashed{\mathcal{D}}\varrho^{m}-\frac{1}{12}T^{mn}_{\underline{\mu\nu \sigma}}\Gamma^{\underline{\mu\nu \sigma}}\varrho_{n}& = & 0,\nonumber
\end{eqnarray}
where in formulas \eqref{6dtensorvar} and \eqref{6dtensoreom}, the covariant derivatives and covariant d'Alembertians are defined as
\begin{eqnarray}
\mathcal{D}_{\underline{\mu}}\varrho^{m}  & = & \left(\partial_{\underline{\mu}}-\frac{5}{2}b_{\underline{\mu}}+\frac{1}{4}\omega^{\underline{ab}}_{\underline{\mu}}\Gamma_{\underline{ab}}\right)\varrho^{m}-\frac{1}{2}V_{\underline{\mu}n}^{\phantom{[}m}\varrho^{n},\nonumber \\
\mathcal{D}_{\underline{\mu}}\Phi^{mn} & = & \left(\partial_{\underline{\mu}}-2b_{\underline{\mu}}\right)\Phi^{mn}+V_{\underline{\mu} r}^{[m}\Phi^{n]r}, \label{6dtensorcovards}\\
\mathcal{D}^{2}\Phi^{mn} & = & \left(\phantom{\int}\hspace{-.15in}\partial^{\underline{a}}-3b^{\underline{a}}+\omega^{\underline{ba}}_{\underline{b}}\right)\mathcal{D}_{\underline{a}}\Phi^{mn}+V_{\phantom{aa}r}^{\underline{a}[m}\mathcal{D}_{\underline{a}}\Phi^{n]r}-4f_{\underline{a}}^{\underline{a}}\Phi^{mn}. \nonumber
\end{eqnarray}
A significant feature of these equations is that they imply that in a general supergravity background, the field strength $H=dB$ is not self-dual, but rather its anti-self-dual part is fixed by \eqref{6dtensoreom} in terms of the dynamical scalar fields $\Phi$.

\subsection{Reduction to $5d$ Abelian Yang-Mills}
We now move on to describe the reduction of tensor multiplet to five dimensions in a general supergravity background.  The $6d$ fields decompose to $5d$ fields as follows
\begin{equation}
\varrho^{m\underline{\beta}}\rightarrow \frac{\alpha}{4}\left(\begin{array}{c} 0 \\ i \rho^{m \beta}\end{array}\right), \hspace{.5in}\Phi^{mn}\rightarrow \alpha \varphi^{mn}, \hspace{.5in} B_{\underline{ab}}\rightarrow B_{c5}\equiv \alpha A_{c}.
\end{equation}
The duality constraints on $H$ imply that the independent five-dimensional degrees of freedom descending from $B$ comprise a five-dimensional gauge field $A$.  The rescaling of the gauge field by the dilation ensures that $A$ has its canonical scaling dimension.  The additional rescalings of the scalars and fermions are chosen so that the final supersymmetric action takes a simple form with the $5d$ Yang-Mills coupling appearing as a prefactor to the Lagrangian.

We now reduce the equations of motion \eqref{6dtensoreom} to relations on the $5d$ fields.  To express the results, it is convenient to define the following functions of the supergravity fields
\begin{eqnarray}
(M_{\varphi})^{rs}_{mn} & = &  \left[\left(\frac{1}{20\alpha^{2}}G_{ab}G^{ab}-\frac{R}{5}\right)\delta^{r}_{m}\delta^{s}_{n}+\frac{1}{2}\left(S^{r}_{[m}S^{s}_{n]}-S^{s}_{t}S^{t}_{[m}\delta^{r}_{n]}\right)-\frac{1}{15}D^{rs}_{mn}-T^{ab}_{mn}T^{rs}_{ab}\right] , \nonumber\\
(M_{\rho})^{mn \alpha}_{\phantom{mn}\beta}  & = &\left[\frac{1}{2}S^{mn}\delta^{\alpha}_{\beta}+\frac{1}{8\alpha}G_{ab}\left(\Gamma^{ab}\right)^{\alpha}_{\beta}\Omega^{mn}-\frac{1}{2}T^{mn}_{ab}\left(\Gamma^{ab}\right)^{\alpha}_{\beta}\right].  \label{sugramass}
\end{eqnarray}
Then, the equations of motion for the $5d$ fields take the form of free field equations with the operators $M_{\varphi}$ and $M_{\rho}$ appearing as mass terms for the scalars and fermions respectively.
\begin{eqnarray}
d\left(\phantom{\int}\hspace{-.15in}\alpha * F -\alpha *\varphi_{mn}T^{mn} \right) +G\wedge F& = & 0, \nonumber \\
\mathcal{D}^{2}\varphi_{mn}+2F_{ab}T^{ab}_{mn}+ (M_{\varphi})^{rs}_{mn} \varphi_{rs} & = & 0, \label{eomasym}\\
i\slashed{D}^{\alpha}_{\beta}\rho^{m\beta}+(M_{\rho})^{mn \alpha}_{\phantom{mn}\beta} \rho^{\beta}_{n}& = & 0. \nonumber
\end{eqnarray}
In the above, $F=dA$ indicates the five-dimensional gauge field strength, $*$ the five-dimensional Hodge dual, and the various covariant derivatives, covariant d'Alembertians and curvatures are given by 
\begin{eqnarray}
\mathcal{D}_{\mu}\rho^{m} & = & \left(\partial_{\mu}-\frac{3}{2}\partial_{\mu}\log(\alpha)+\frac{1}{4}\omega^{bc}_{\mu}\Gamma_{bc}\right)\rho^{m}-\frac{1}{2}V_{\mu n}^{m}\rho^{n},\nonumber \\
\mathcal{D}_{\mu}\varphi_{mn} & = & \left(\phantom{\int}\hspace{-.15in}\partial_{\mu}-\partial_{\mu}\log(\alpha)\right)\varphi_{mn}-V_{\mu[m}^{\phantom{\mu}r}\varphi_{n]r}, \label{abeliancovardmat}\\
\mathcal{D}^{2}\varphi_{mn} & = & \left(\phantom{\int}\hspace{-.15in}\partial^{a}-2\partial^{a}\log(\alpha)+\omega^{ba}_{b}\right)\mathcal{D}_{a}\varphi_{mn}-V_{a[m}^{\phantom{a}r}\mathcal{D}^{a}\varphi_{n]r}, \nonumber \\
R & = & e^{\mu}_{a}e^{\nu}_{b}\left(\phantom{\int}\hspace{-.15in}2\partial_{[\mu}\omega_{\nu]}^{ab}+2\omega_{[\mu}^{ac}\omega_{\nu]c}^{\phantom{\nu]cc}b}\right). \nonumber
\end{eqnarray}
In the special case where the dilaton is constant, $R$ is the Ricci scalar curvature of the five-dimensional metric. 

 In general, the derivative operator $\mathcal{D}_{\mu}$ is covariant with respect to the Levi-Cevita connection, the $sp(4)$ R-symmetry and the connection $\partial_{\mu}\alpha$ which acts as a gauge field for local rescalings.  One consequence of this, useful in many integration by parts manipulations, is that the dilation field $\alpha$ is covariantly constant
\begin{equation}
\mathcal{D}_{\mu}\alpha =0.
\end{equation}

Finally, the supersymmetry transformations reduce to
\begin{eqnarray}
\delta A_{c} & = &- \frac{i}{4}\epsilon_{m}\Gamma_{c}\rho^{m}, \nonumber \\
\delta \varphi^{mn} & = &- \epsilon^{[m}\rho^{n]}-\frac{1}{4}\Omega^{mn}\epsilon^{r}\rho_{r}, \label{5dsymat}\\
\delta \rho^{m} & = & \left(\phantom{\int}\hspace{-.15in}S^{[m}_{s}\varphi^{n]s}\Omega_{rn}-2\varphi^{mn}S_{nr}-i\slashed{\mathcal{D}}\varphi^{mn}\Omega_{rn}\right)\epsilon^{r} \nonumber \\
& + & \frac{1}{4}\left(\phantom{\int}\hspace{-.15in} 2F^{ab}\delta^{m}_{r}-\varphi^{ns}T_{ns}^{ab}\delta^{m}_{r}-4\varphi^{mn}T_{nr}^{ab}-\frac{2}{\alpha}\varphi^{mn}G^{ab}\Omega_{rn}\right)\Gamma_{ab}\epsilon^{r}.\nonumber
\end{eqnarray}

Although these supersymmetry variations as well as the equations of motion \eqref{eomasym} are a direct consequence of their six-dimensional counterparts, there is an important simplification in five dimensions: the equations of motion may be derived from an action invariant under the supersymmetry transformations stated in \eqref{5dsymat}.  This action may be constructed by integrating the equations of motion.  Define
\begin{eqnarray}
S_{A} & = &-\frac{1}{8\pi^{2}}\int \ \left[\phantom{\int}\hspace{-.15in}\alpha F\wedge *F+G\wedge A \wedge F\right], \nonumber\\
S_{\varphi} & = &\frac{1}{32\pi^{2}} \int d^{5}x \sqrt{|g|} \ \alpha \varphi^{mn}\left(\phantom{\int}\hspace{-.15in} \mathcal{D}^{2}\varphi_{mn}+4F_{ab}T^{ab}_{mn}+(M_{\varphi})^{rs}_{mn} \varphi_{rs}\right), \label{5dabelianaction}\\
S_{\rho} & = &-\frac{1}{32\pi^{2}} \int d^{5}x  \sqrt{|g|}\ \alpha \rho_{m\gamma}\left(\phantom{\int}\hspace{-.15in} i\slashed{\mathcal{D}}^{\gamma}_{\beta}\rho^{m\beta}+(M_{\rho})^{mn \gamma}_{\phantom{mn}\beta} \rho^{\beta}_{n} \right). \nonumber
\end{eqnarray}
The total action is then 
\begin{equation}
S=S_{A}+S_{\varphi}+S_{\rho}. 
\end{equation}
It is invariant under the supersymmetry transformations \eqref{5dsymat} provided that the background fields and spinor parameters are supersymmetric as defined by the vanishing of equations \eqref{5dsusyvar}.
\subsection{Non-Abelian Extension}
\label{nonabext}
We now seek to generalize the abelian action derived in previous sections, to an action for non-abelian five-dimensional Yang-Mills coupled to off-shell supergravity background fields.  Unlike the abelian theory, the non-abelian six-dimensional $(2,0)$ superconformal theory admits no known formulation in terms of fields and equations of motion.  Thus our strategy is to construct the non-abelian generalization of the five-dimensional action \eqref{5dabelianaction} directly.  Such a generalization is possible as a consequence of the following basic logic.
\begin{itemize}
\item The background supergravity multiplet, as well as the supersymmetry conditions \eqref{5dsusyvar} on spinor parameters $\epsilon$ follow from the superconformal algebra in six dimensions.  Thus they receive no modifications when passing from the abelian to non-abelian action.
\item The non-abelian action in flat space is known.  It is constructed by covariantizing the abelian action \eqref{5dabelianaction} with respect to non-abelian gauge invariance, adding Yukawa couplings and a quartic scalar potential, and finally modifying the supersymmetry variation of the fermions to include a commutator of bosons.  
\item The supersymmetry transformations must be consistent with the Weyl rescaling symmetry inherited from the six-dimensional theory.  The most general expression is the abelian transformation \eqref{5dsymat} together with the non-abelian commutator term that appears in flat space.  No further corrections to the supersymmetry transformations are possible.
\item The non-abelian action in the presence of general supergravity backgrounds is constrained by gauge invariance, R-symmetry invariance, dilation invariance, and Lorentz invariance.  Further, the action may be consistently truncated to terms of total Weyl weight five.  This yields a single allowed truly non-abelian term which vanishes in the trivial supergravity background.  This term takes the form of a cubic coupling of scalar fields 
\begin{equation}
\alpha S_{mn} \mathrm{Tr}\left(\phantom{\int}\hspace{-.16in}\varphi^{mr}[\varphi^{ns},\varphi_{rs}]\right).
\end{equation}
We determine the coefficient of this term by demanding supersymmetry of the action.  We find that this coefficient is non-zero.
\end{itemize}
In the remainder of this section we apply the above reasoning to determine the complete non-abelian action coupled to general off-shell supergravity backgrounds.  The final form of the supersymmetry transformations and action are stated in \eqref{5dsynonmat} and \eqref{5dnonabelianaction}.

To begin, let $\mathfrak{g}$ indicate a simple compact Lie algebra.  We use conventions such that a matrix $X\in \mathfrak{g}$ when $\exp(t X)$ is in the associated Lie group for all real $t$.  The Lie algebra is closed under commutators and equipped with a negative definite invariant bilinear form which we indicate by $\mathrm{Tr}$.  Thus, in the case $su(n)$-type Lie algebras, the generators are antihermitian, and the bilinear form is the trace in any fixed representation.

All matter fields in our theory transform in the adjoint representation, thus from now on the symbols $\varphi, \rho$, and $A_{\mu}$ will denote Lie algebra valued fields.  If $\varsigma$ indicates a Lie algebra valued gauge parameter the associated variation of the fields is given by  
\begin{eqnarray}
\varphi_{mn} &\rightarrow &\varphi_{mn}+[\varsigma,\varphi_{mn}], \nonumber \\
\rho_{m}^{\alpha}& \rightarrow & \rho_{m}^{\alpha}+[\varsigma,\rho_{m}^{\alpha}], \\
A_{\mu}& \rightarrow & A_{\mu}+[\varsigma, A_{\mu}]-\partial_{\mu}\varsigma. \nonumber
\end{eqnarray}
To account for these transformations, we modify the definitions of field strengths and covariant derivatives from their abelian form \eqref{abeliancovardmat} to
\begin{eqnarray}
\mathcal{D}_{\mu}\rho^{m} & = & \left(\partial_{\mu}-\frac{3}{2}\partial_{\mu}\log(\alpha)+\frac{1}{4}\omega^{bc}_{\mu}\Gamma_{bc}\right)\rho^{m}-\frac{1}{2}V_{\mu n}^{m}\rho^{n}+[A_{\mu},\rho^{m}],\nonumber \\
\mathcal{D}_{\mu}\varphi_{mn} & = & \left(\phantom{\int}\hspace{-.15in}\partial_{\mu}-\partial_{\mu}\log(\alpha)\right)\varphi_{mn}-V_{\mu[m}^{\phantom{\mu}r}\varphi_{n]r}+[A_{\mu},\varphi_{mn}], \label{nonabeliancovardmat}\\
F_{\mu\nu} & = & \partial_{\mu}A_{\nu}-\partial_{\nu}A_{\mu}+[A_{\mu},A_{\nu}]. \nonumber
\end{eqnarray}
The non-abelian modifications are the commutators appearing on the right-hand-side of the above.  Similarly, we modify the supersymmetry transformations of the fields by including gauge covariant terms as well as a commutator of scalars in the variation of the fermions $\rho^{m}$
\begin{eqnarray}
\delta A_{c} & = &- \frac{i}{4}\epsilon_{m}\Gamma_{c}\rho^{m}, \nonumber \\
\delta \varphi^{mn} & = &- \epsilon^{[m}\rho^{n]}-\frac{1}{4}\Omega^{mn}\epsilon^{r}\rho_{r}, \label{5dsynonmat}\\
\delta \rho^{m} & = & \left(\phantom{\int}\hspace{-.15in}S^{[m}_{s}\varphi^{n]s}\Omega_{rn}-2\varphi^{mn}S_{nr}-i\slashed{\mathcal{D}}\varphi^{mn}\Omega_{rn}\right)\epsilon^{r} -\frac{1}{2}\Omega_{nr}[\varphi^{mn}, \varphi^{rs}]\epsilon_{s}\nonumber \\
& + & \frac{1}{4}\left(\phantom{\int}\hspace{-.15in} 2F^{ab}\delta^{m}_{r}-\varphi^{ns}T_{ns}^{ab}\delta^{m}_{r}-4\varphi^{mn}T_{nr}^{ab}-\frac{2}{\alpha}\varphi^{mn}G^{ab}\Omega_{rn}\right)\Gamma_{ab}\epsilon^{r}.\nonumber
\end{eqnarray}
All terms in the above now have their standard non-abelian definitions.  In the flat space limit, \eqref{5dsynonmat} reduces to the standard supersymmetry transformations on the non-abelain fields.  To justify the above transformations in the general supergravity background we note that no further non-abelian terms are permitted by Weyl invariance.  Any non-abelian supergravity correction to the supersymmetry transformations must involve a commutator of Yang-Mills fields times a product of supergravity background fields and the supersymmetry parameter $\epsilon$ .  However, the tuple $(A,\varphi, \rho,\epsilon)$ has scaling dimensions $(1,1,3/2,-1/2)$, and all supergravity fields have positive scaling dimensions.  Thus, the most general non-abelian correction to the algebra appears in \eqref{5dsynonmat}.

Next we come to the action.  The abelian kinetic terms are covariantized.  In addition, a Yukawa and quartic scalar interaction are added as dictated by the non-abelian Lagrangian in flat space.  Finally, a non-abelian cubic scalar interaction is added which vanishes in flat space.  The non-abelian interaction terms are written in the action $S_{int}$ below.
\begin{eqnarray}
S_{A} & = &\frac{1}{8\pi^{2}}\int \mathrm{Tr} \left(\phantom{\int}\hspace{-.15in}\alpha F\wedge *F+C\wedge F \wedge F\right), \nonumber\\
S_{\varphi} & = &\frac{1}{32\pi^{2}} \int d^{5}x \sqrt{|g|} \ \alpha \mathrm{Tr}\left(\phantom{\int}\hspace{-.15in} \mathcal{D}_{a}\varphi^{mn}\mathcal{D}^{a}\varphi_{mn}-4\varphi^{mn}F_{ab}T^{ab}_{mn}-\varphi^{mn}(M_{\varphi})^{rs}_{mn} \varphi_{rs}\right),  \nonumber\\
S_{\rho} & = &\frac{1}{32\pi^{2}} \int d^{5}x  \sqrt{|g|}\ \alpha\mathrm{Tr}\left(\phantom{\int}\hspace{-.15in}  \rho_{m\gamma}i\slashed{\mathcal{D}}^{\gamma}_{\beta}\rho^{m\beta}+ \rho_{m\gamma}(M_{\rho})^{mn \gamma}_{\phantom{mn}\beta} \rho^{\beta}_{n} \right). \label{5dnonabelianaction}\\
S_{int} & = &\frac{1}{32\pi^{2}} \int d^{5}x  \sqrt{|g|}\ \alpha \mathrm{Tr}\left(\phantom{\int}\hspace{-.15in}\rho_{m\alpha}[\varphi^{mn},\rho_{n}^{\alpha}]-\frac{1}{4}[\varphi_{mn},\varphi^{nr}][\varphi_{rs},\varphi^{sm}]-\frac{2}{3}S_{mn} \varphi^{mr}[\varphi^{ns},\varphi_{rs}]\right)\nonumber
\end{eqnarray}
In the above, the supergravity induced mass terms $M_{\varphi}$ and $M_{\rho}$ are defined, as in the case of the abelian theory by \eqref{sugramass}.  In $S_{int},$ the coefficient of the cubic scalar interaction proportional to the field $S_{mn}$ is fixed by demanding supersymmetry in the presence of backgrounds where this field is activated subject to the constraints determined by \eqref{5dsusyvar}.  The action \eqref{5dnonabelianaction} is our final result.

\section*{Acknowledgements} 
The work of C.C. is support by a Junior Fellowship at the Harvard Society of Fellows.  The work of D.J. is supported by the Fundamental Laws Initiative Fund at Harvard University, and the National Science Foundation Grant No. 1066293.

\appendix
\section{Conventions}
\label{convent}
Here we describe our conventions indices, differential forms and Clifford algebra.

We adopt the following index labeling conventions:
\begin{itemize}
\item Greek indices from the first half of the alphabet (e.g. $\alpha, \beta$ etc.) indicate spinor components.
\item Greek indices from the last half of the alphabet (e.g. $\mu, \nu$ etc.) indicate coordinate basis component on the tangent bundle.
\item Latin indices from the first half of the alphabet (e.g. $a, b$ etc.) indicate orthonormal frame components on the tangent bundle.
\item Latin indices from the last half of the alphabet (e.g $m, n$) indicate the fundamental representation of $sp(4)$.
\item When necessary, any six-dimensional index is underlined to distinguish it from its five-dimensional descendants.
\end{itemize}
\subsection{Differential Forms}
Our conventions for differential forms are standard and are recorded here for convenience.  Let $\Xi$ indicate a $k$-form in $n$ total dimensions.  The form $\Xi$ may be expanded in components as
\begin{equation}
\Xi=\left(\frac{1}{k!}\right)\Xi_{\mu_{1}\mu_{2}\cdots \mu_{k}}dx^{\mu_{1}}\wedge dx^{\mu_{2}}\wedge \cdots \wedge dx^{\mu_{k}},
\end{equation}
where the tensor $\Xi_{\mu_{1}\mu_{2}\cdots \mu_{k}}$ is totally antisymmetric.
The exterior derivative of $\Xi$ is written in explicit coordinates as
\begin{equation}
(d \Xi)_{\mu_{1}\mu_{2}\cdots \mu_{k+1}}=(k+1)\partial_{[\mu_{1}}\Xi_{\mu_{2}\mu_{3}\cdots \mu_{k+1}]}.
\end{equation}
If $\Upsilon$ indicates a $p$-form, then the wedge product $\Upsilon \wedge \Xi$ is a $k+p$-form with components
\begin{equation}
(\Upsilon \wedge \Xi)_{\mu_{1}\mu_{2}\cdots \mu_{k+p}}=\frac{(p+k)!}{(p)!(k)!}\Upsilon_{[\mu_{1}\cdots \mu_{p}}\Xi_{\mu_{p+1}\cdots \mu_{p+k}]}.
\end{equation}

Finally, we also have need of the Hodge $*$ operation.  Let $\varepsilon_{a_{1}a_{2}\cdots a_{n}}$ be the totally antisymmetric Levi-Cevita symbol.  As written in frame indices it satisfies
\begin{equation}
\varepsilon_{0 1 \cdots n-1}=1.
\end{equation}
As with all tensors, its frame indices may be converted to tangent indices by making use of the orthonormal frame, and may be raised an lowered using the metric tensor.  From this tensor we may define the $*$ map as
\begin{equation}
(*\Xi)_{\mu_{1}\cdots \mu_{n-k}}=\frac{1}{k!}\varepsilon_{\mu_{1}\cdots \mu_{n-k}\nu_{1}\cdots \nu_{k}}\Xi^{\nu_{1}\cdots \nu_{k}}.
\end{equation}

\subsection{$6d$ Spinors}
We work throughout in the a mostly plus signature $(- + \cdots +)$ for a Lorentzian metric.   The six-dimensional Clifford algebra is defined by
\begin{equation}
\{\Gamma_{\underline{a}},\Gamma_{\underline{b}}\}^{\underline{\alpha}}_{\underline{\beta}}=2\eta_{\underline{ab}}\delta^{\underline{\alpha}}_{\underline{\beta}}.
\end{equation}
The spinor index $\underline{\alpha}$ ranges from one to eight.  The matrices $\Gamma_{\underline{a}}$ associated to spatial directions may be taken to be hermitian, while $\Gamma_{\underline{0}}$ is antihermitian.  Suppressed spinor indices are contracted using the natural pairing between a representation and its dual as discussed further in the following.

The generators of Lorentz transformations acting on spinors are constructed from the Clifford algebra in the standard fashion
\begin{equation}
M_{\underline{ab}}=\frac{i}{4}\left[\Gamma_{\underline{a}},\Gamma_{\underline{b}}\right].
\end{equation}
As in all even dimensions, the Dirac spinor representation is reducible into two Weyl spinors.  Introduce a chirality matrix
\begin{equation}
\Gamma=-\Gamma_{\underline{0}}\Gamma_{\underline{1}}\Gamma_{\underline{2}}\Gamma_{\underline{3}}\Gamma_{\underline{4}}\Gamma_{\underline{5}}, \hspace{.5in} \{\Gamma,\Gamma_{\underline{a}}\}=0, \hspace{.5in} (\Gamma^{2})^{\underline{\alpha}}_{\underline{\beta}}=\delta^{\underline{\alpha}}_{\underline{\beta}}.
\end{equation}
The positive and negative chirality Weyl spinors, $W_{\pm}$, are then the positive and negative eigenspaces of $\Gamma$ respectively.

The Dirac spinor is isomorphic as a representation to both its complex conjugate and dual.  However, the Weyl representations are self-conjugate but dual to each other
\begin{equation}
\overline{W}_{\pm}\cong W_{\pm}, \hspace{.5in} \check{W}_{\pm}\cong W_{\mp}.
\end{equation}
These statements mean in particular that we may find matrices $\underline{B}$ and $\underline{C}$ facilitating the previously stated isomorphisms for Dirac spinor representations
\begin{equation}
\underline{B}\Gamma_{\underline{a}} \underline{B}^{-1}=\Gamma^{*}_{\underline{a}}, \hspace{.5in}\underline{C}\Gamma_{\underline{a}} \underline{C}^{-1}=-\Gamma_{\underline{a}}^{T}.
\end{equation}

Here $\underline{C}$ is the isomorphism (often called the `charge conjugation matrix') between the Dirac spinor and its dual and $\underline{C}^{-1}$ as the inverse isomorphism.  With the convention that raised indices represent Dirac spinors, lowered indices their duals, and hence the natural pairing between them given by contraction of upper and lower indices, we may use the charge conjugation matrix to raise and lower spinor indices.
\begin{equation}
\psi_{\underline{\alpha}}=\psi^{\underline{\beta}}\underline{C}_{\underline{ \beta \alpha}}, \hspace{.5in} \psi^{\underline{\alpha}}=\underline{C}^{\underline{\alpha\beta}}\psi_{\underline{\beta}} \label{cdual}
\end{equation}
Similarly, $\underline{B}$ is the isomorphism between the Dirac spinor and its complex conjugate.  As the Weyl spinors are also self-conjugate in this dimension, the restriction of $\underline{B}$ to the eigenspaces of $\Gamma$ provides an isomorphism between the Weyl spinors and their complex conjugates.

From the Clifford algebra we may construct a standard set of operators on spinors, which transform as antisymmetric tensors.  We denote these as multi-index gamma matrices defined as antisymmetric products
\begin{equation}
\Gamma^{\underline{a_{1}a_{2}\cdots a_{n}}}=\frac{1}{n!}\sum_{\sigma\in S_{n}}(-1)^{|\sigma|}\Gamma^{\underline{a}_{\sigma(1)}}\Gamma^{\underline{a}_{\sigma(2)}}\cdots \Gamma^{\underline{a}_{\sigma(n)}}.
\end{equation}
These matrices allow us to construct fermion bilinears of the form, $\chi \Gamma^{\underline{a_{1}a_{2}\cdots a_{n}}} \psi$, and in such expressions we set the convention that whenever contracted spinor indices are omitted they are summed southwest to northeast. 
\begin{equation}
\chi \psi \equiv \chi_{\underline{\alpha}}\psi^{\underline{\alpha}}.
\end{equation}
Finally, we note that multiplication of the above multicomponent gamma matrices by the chirality matrix $\Gamma$ allows us to relate those with $n$ indices to those with $6-n$.  Of particular importance is the relation
\begin{equation}
\Gamma_{\underline{a_{1}a_{2}a_{3}}}\Gamma=-\frac{1}{6}\varepsilon_{\underline{a_{1}a_{2}a_{3}a_{4}a_{5}a_{6}}}\Gamma^{\underline{a_{4}a_{5}a_{6}}},
\end{equation}
where the Levi-Cevita symbol is totally antisymmetric with $\varepsilon_{\underline{012345}}=1$.
\subsubsection{The Symplectic Majorana-Weyl  Condition}
\label{6dsymplecticmajorana}
All six-dimensional spinors appearing in this paper also transform in non-trivial representations of the R-symmetry $sp(4)$.  Raised Latin indices from the last half of the alphabet denote components of the fundamental four-dimensional representation $sp(4)$.  Since $sp(4)$ is a simple Lie algebra its finite dimensional representations are unitary and hence the complex conjugate and dual of any finite dimensional representations are naturally isomorphic.  In the case of the fundamental, there is a further isomorphism with the representation itself
\begin{equation}
\mathbf{4} \cong \check{\mathbf{4}} \cong \overline{\mathbf{4}}.
\end{equation}
This isomorphism exists because of the two index antisymmetric tensor, $\Omega$, preserved by all elements of $sp(4)$.  In explicit notation we let a lowered $sp(4)$ index denote an element of $\check{\mathbf{4}}$ and the natural pairing between a representation and its dual as contraction of upper and lower indices.  Then, $\Omega$ is a two index object which may be used to raise an lower $sp(4)$ indices
\begin{equation}
\lambda_{m}=\lambda^{n}\Omega_{nm}, \hspace{.5in} \lambda^{m}=\Omega^{mn}\lambda_{n}, \hspace{.5in} \Omega_{mn}=-\Omega_{nm}.
\end{equation}
In general, we set the convention that whenever contracted $sp(4)$ indices are omitted, they are summed northwest to southeast.

Let us now consider a Weyl spinor also transforming in the $\mathbf{4}$ of the symplectic group, $\psi ^{\underline{\alpha}m}$.  On this representation there is a reality condition available which in our explicit index notation takes the form
\begin{equation}
\left(\psi^{\underline{\alpha}}_{m}\right)^{*}=\underline{B}_{\underline{\beta}}^{\underline{\alpha}}\psi^{\underline{\beta} m}. \label{smw1}
\end{equation}
Weyl spinors satisfying this condition are said to be symplectic Majorana-Weyl.  They comprise a representation of $so(1,5)\times sp(4)$ of real dimension 16.  In practice the above is frequently rewritten in terms of the charge conjugation matrix.  Define the Dirac conjugate spinor $\overline{\psi}^{m}$ as
\begin{equation}
\overline{\psi}^{m}_{\underline{\beta}}=\left(\psi_{m}^{\underline{\alpha}}\right)^{*}\Gamma_{\underline{0}\underline{\beta}}^{\underline{\alpha}}. \label{psibarder}
\end{equation}
Then the symplectic Majorana constraint takes the simple form
\begin{equation}
\overline{\psi}^{m}_{\underline{\beta}} = \psi^{ \underline{\alpha}m}C_{\underline{\alpha \beta}}=\psi^{m}_{\underline{\beta}}. \label{smw2}
\end{equation}
\subsection{$5d$ Spinors}
\label{5dcliffcon}
We now turn our attention to five-dimensional spinors.  The Clifford algebra takes the usual form
\begin{equation}
\{\Gamma_{a},\Gamma_{b}\}^{\alpha}_{\beta}=2\eta_{ab}\delta^{\alpha}_{\beta}.
\end{equation}
However, now the spinor index $\alpha$ ranges from one to four and comprises the irreducible Dirac spinor of $so(1,4)$.

As in six dimensions we construct antisymmetric tensors valued in operators on spinors by taking antisymmetrized products.
\begin{equation}
\Gamma^{a_{1}a_{2}\cdots a_{n}}=\frac{1}{n!}\sum_{\sigma\in S_{n}}(-1)^{|\sigma|}\Gamma^{a_{\sigma(1)}}\Gamma^{a_{\sigma(2)}}\cdots \Gamma^{a_{\sigma(n)}}.
\end{equation}
In particular, the top form valued matrix above is
\begin{equation}
\left(\Gamma_{a_{1}a_{2}a_{3}a_{4}a_{5}}\right)^{\alpha}_{\beta}=i \varepsilon_{a_{1}a_{2}a_{3}a_{4}a_{5}}\delta^{\alpha}_{\beta},
\end{equation}
and our convention is that $\varepsilon_{01234}=1$.

Finally, let us discuss the reality properties of five-dimensional spinors.  The Dirac spinor is isomorphic to its complex conjugate and dual representations.  Hence we may again find matrices $B$ and $C$ such that
\begin{equation}
B\Gamma_{a} B^{-1}=-\overline{\Gamma}_{a}, \hspace{.5in}C\Gamma_{a} C^{-1}=\Gamma_{a}^{T}. \label{5dbdef}
\end{equation}

All five-dimensional spinors appearing in this paper transform in non-trivial representations of the R-symmetry $sp(4)$.  Our index conventions for these transformations are identical to section \ref{6dsymplecticmajorana}.   On any Dirac spinor transforming in the $\mathbf{4}$ of $sp(4)$ we may impose a symplectic Majorana condition
\begin{equation}
\left(\psi^{\alpha}_{m}\right)^{*}=B_{\beta}^{\alpha}\psi^{\beta m} \Leftrightarrow \overline{\psi}^{m}_{\alpha}=\psi^{\beta m}C_{\beta \alpha}=\psi^{m}_{\alpha}.
\end{equation}
Symplectic Majorana spinors comprise a representation of $so(1,4)\times sp(4)$ of real dimension 16.
\subsection{Reducing $6d$ Spinors to $5d$}
\label{6d5dspinors}
To relate spinors in five and six dimensions we first specify the relations on the associated Clifford algebras.  Let $\sigma_{1}, \sigma_{2}, \sigma_{3}$ indicate the Pauli matrices
\begin{equation}
\sigma_{1}=\left(\begin{array}{cc} 0 & 1 \\ 1 & 0\end{array}\right), \hspace{.5in} \sigma_{2}=\left(\begin{array}{cc} 0 & -i \\ i & 0\end{array}\right), \hspace{.5in} \sigma_{3}=\left(\begin{array}{cc} 1 & 0 \\ 0 & -1\end{array}\right).
\end{equation}
Then the six-dimensional gamma matrices may be constructed from those in five dimensions as
\begin{equation}
\Gamma_{\underline{a}}=\begin{cases} \Gamma_{a}\otimes \sigma_{1} & \mathrm{if} \ \underline{a} \leq4, \\
\mathbf{1}_{4}\otimes \sigma_{2} & \mathrm{if} \ \underline{a}=5.
\end{cases}
\end{equation}
In particular, we learn from this that the six-dimensional chirality matrix and charge conjugation matrix are
\begin{equation}
\Gamma= \mathbf{1}_{4}\otimes \sigma_{3}, \hspace{.5in}\underline{C}=C\otimes i\sigma_{2}. \label{chiral65}
\end{equation}
We may now reduce six-dimensional spinors to five-dimensional representations.  We write a six-dimensional Dirac spinor transforming in the $\mathbf{4}$ as a doublet compatible with the tensor product description given above
\begin{equation}
\psi^{m}=\left(\begin{array}{c} \psi_{+}^{m} \\ i\psi_{-}^{m}\end{array}\right).
\end{equation}
Each component $\psi^{m}_{\pm}$ transforms as a spinor of $so(1,4)$.  According to \eqref{chiral65}, the subscript $\pm$ indicates the six-dimensional chirality of the spinor, and $\psi^{m}$ is Majorana in six dimensions if and only if $\psi_{\pm}^{m}$ are Majorana in five dimensions.

\bibliography{Arxiv.bbl}{}

\providecommand{\href}[2]{#2}\begingroup\raggedright\begin{thebibliography}{10}

\bibitem{Bergshoeff:1999db}
E.~Bergshoeff, E.~Sezgin, and A.~Van~Proeyen, ``{(2,0) tensor multiplets and
  conformal supergravity in D = 6},''
  \href{http://dx.doi.org/10.1088/0264-9381/16/10/311}{{\em Class.Quant.Grav.}
  {\bfseries 16} (1999) 3193--3206},
\href{http://arxiv.org/abs/hep-th/9904085}{{\ttfamily arXiv:hep-th/9904085
  [hep-th]}}.

\bibitem{Riccioni:1997np}
F.~Riccioni, ``{Tensor multiplets in six-dimensional (2,0) supergravity},''
  \href{http://dx.doi.org/10.1016/S0370-2693(98)00070-7}{{\em Phys.Lett.}
  {\bfseries B422} (1998) 126--134},
\href{http://arxiv.org/abs/hep-th/9712176}{{\ttfamily arXiv:hep-th/9712176
  [hep-th]}}.

\bibitem{Kugo:2000hn}
T.~Kugo and K.~Ohashi, ``{Supergravity tensor calculus in 5-D from 6-D},''
  \href{http://dx.doi.org/10.1143/PTP.104.835}{{\em Prog.Theor.Phys.}
  {\bfseries 104} (2000) 835--865},
\href{http://arxiv.org/abs/hep-ph/0006231}{{\ttfamily arXiv:hep-ph/0006231
  [hep-ph]}}.

\bibitem{Kugo:2000af}
T.~Kugo and K.~Ohashi, ``{Off-shell D = 5 supergravity coupled to matter
  Yang-Mills system},'' \href{http://dx.doi.org/10.1143/PTP.105.323}{{\em
  Prog.Theor.Phys.} {\bfseries 105} (2001) 323--353},
\href{http://arxiv.org/abs/hep-ph/0010288}{{\ttfamily arXiv:hep-ph/0010288
  [hep-ph]}}.

\bibitem{Fujita:2001kv}
T.~Fujita and K.~Ohashi, ``{Superconformal tensor calculus in
  five-dimensions},'' \href{http://dx.doi.org/10.1143/PTP.106.221}{{\em
  Prog.Theor.Phys.} {\bfseries 106} (2001) 221--247},
\href{http://arxiv.org/abs/hep-th/0104130}{{\ttfamily arXiv:hep-th/0104130
  [hep-th]}}.

\bibitem{Bergshoeff:1985mz}
E.~Bergshoeff, E.~Sezgin, and A.~Van~Proeyen, ``{SUPERCONFORMAL TENSOR CALCULUS
  AND MATTER COUPLINGS IN SIX-DIMENSIONS},''
\href{http://dx.doi.org/10.1016/0550-3213(86)90503-1}{{\em Nucl.Phys.}
  {\bfseries B264} (1986) 653}.

\bibitem{Coomans:2011ih}
F.~Coomans and A.~Van~Proeyen, ``{Off-shell N=(1,0), D=6 supergravity from
  superconformal methods},'' \href{http://dx.doi.org/10.1007/JHEP02(2011)049,
  10.1007/JHEP01(2012)119}{{\em JHEP} {\bfseries 1102} (2011) 049},
\href{http://arxiv.org/abs/1101.2403}{{\ttfamily arXiv:1101.2403 [hep-th]}}.

\bibitem{Linander:2011jy}
H.~Linander and F.~Ohlsson, ``{(2,0) theory on circle fibrations},''
  \href{http://dx.doi.org/10.1007/JHEP01(2012)159}{{\em JHEP} {\bfseries 1201}
  (2012) 159},
\href{http://arxiv.org/abs/1111.6045}{{\ttfamily arXiv:1111.6045 [hep-th]}}.

\bibitem{Pestun:2007rz}
V.~Pestun, ``{Localization of gauge theory on a four-sphere and supersymmetric
  Wilson loops},'' \href{http://dx.doi.org/10.1007/s00220-012-1485-0}{{\em
  Commun.Math.Phys.} {\bfseries 313} (2012) 71--129},
\href{http://arxiv.org/abs/0712.2824}{{\ttfamily arXiv:0712.2824 [hep-th]}}.

\bibitem{Kallen:2012cs}
J.~Kallen and M.~Zabzine, ``{Twisted supersymmetric 5D Yang-Mills theory and
  contact geometry},'' \href{http://dx.doi.org/10.1007/JHEP05(2012)125}{{\em
  JHEP} {\bfseries 1205} (2012) 125},
\href{http://arxiv.org/abs/1202.1956}{{\ttfamily arXiv:1202.1956 [hep-th]}}.

\bibitem{Hosomichi:2012ek}
K.~Hosomichi, R.-K. Seong, and S.~Terashima, ``{Supersymmetric Gauge Theories
  on the Five-Sphere},''
  \href{http://dx.doi.org/10.1016/j.nuclphysb.2012.08.007}{{\em Nucl.Phys.}
  {\bfseries B865} (2012) 376--396},
\href{http://arxiv.org/abs/1203.0371}{{\ttfamily arXiv:1203.0371 [hep-th]}}.

\bibitem{Kim:2012ava}
H.-C. Kim and S.~Kim, ``{M5-branes from gauge theories on the 5-sphere},''
\href{http://arxiv.org/abs/1206.6339}{{\ttfamily arXiv:1206.6339 [hep-th]}}.

\bibitem{Imamura:2012xg}
Y.~Imamura, ``{Supersymmetric theories on squashed five-sphere},''
  \href{http://dx.doi.org/10.1093/ptep/pts052}{{\em PTEP} {\bfseries 2013}
  (2013) 013B04},
\href{http://arxiv.org/abs/1209.0561}{{\ttfamily arXiv:1209.0561 [hep-th]}}.

\bibitem{Festuccia:2011ws}
G.~Festuccia and N.~Seiberg, ``{Rigid Supersymmetric Theories in Curved
  Superspace},'' \href{http://dx.doi.org/10.1007/JHEP06(2011)114}{{\em JHEP}
  {\bfseries 1106} (2011) 114},
\href{http://arxiv.org/abs/1105.0689}{{\ttfamily arXiv:1105.0689 [hep-th]}}.

\end{thebibliography}\endgroup
\bibliographystyle{utphys}

\end{document}